\documentclass[epj,nopacs]{svjour}

\usepackage{latexsym}
\usepackage{amsmath,amssymb}
\usepackage{threeparttable}
\usepackage{cite}

\begin{document}

\title{Weak magnetism in chiral quark models}

\author{Tommy Ohlsson\thanks{\email{tommy@theophys.kth.se}}, H{\aa}kan
Snellman\thanks{\email{snell@theophys.kth.se}}}

\mail{Tommy Ohlsson}

\institute{Theoretical Physics, Department of
Physics, Royal Institute of Technology, SE-100 44 Stockholm, Sweden}

\date{Received: 29 June 1999}

\abstract{We discuss symmetry breaking in the weak magnetism form factors for
the semileptonic octet baryon decays.  In the chiral quark model, the
symmetry breaking can be accounted for in the masses and the quark spin
polarizations can take on more general values due to Goldstone boson
depolarization.  Here we clarify some features of the chiral quark model
prediction for the weak magnetism and compare to the corresponding result
of the chiral quark soliton model.}

\authorrunning{T. Ohlsson \and H. Snellman}

\titlerunning{Weak magnetism in chiral quark models}

\maketitle


\section{Introduction}

The weak interaction of the baryon octet has been a source of much
information concerning the dynamics of the octet baryons. Especially
the weak interactions can probe details of the hadronic structure at
low energies.

Already early, the conserved vector current (CVC) hypothesis
\cite{feyn58}
extended to flavor SU(3) symmetry made predictions for the so called
weak magnetism form factors.

In the quark model (QM), these predictions are easily rederived in the same
limit.  Introducing explicit SU(3) symmetry breaking through the masses
of baryons and quarks, the QM will be able to model part of the symmetry
breaking of the weak magnetism form factors \cite{kell74,dono82}.

In the chiral quark model ($\chi$QM) \cite{mano84,eich92,chen95}, it is
also possible to include
modifications of the spin polarizations of the quarks in these form
factors.

In this note we discuss these form factors and compare to the same form
factors derived in the so called chiral quark soliton model
($\chi$QSM) \cite{kimp98}.  The result of the latter model seems to
differ in some
respects with the result of the $\chi$QM \cite{ohls99}. We here show that
when linear
terms in the symmetry breaking are eliminated the two models give the
same predictions.

\section{The weak form factors}

The transition matrix element ${\cal M}_{B \to B' l^- \bar{\nu}_l}$
for the decay $B \to B' + l^- + \bar{\nu}_l$ ($q \to q' +
l^- + \bar{\nu}_l$), is given by
\begin{equation}
{\cal M}_{B \to B' l^- \bar{\nu}_l} = \frac{G}{\sqrt{2}} V_{qq'}
\langle B'(p') \vert J_{\rm weak}^\mu \vert B(p) \rangle L_\mu,
\end{equation}
where $G$ is the Fermi coupling constant, $V_{qq'}$ is the
$qq'$-element of the Cabibbo--Kobayashi--Maskawa mixing matrix,
and $L_{\mu}$ is the leptonic current.

The hadronic weak current is
\begin{equation}
J_{\rm weak}^\mu = J_V^\mu - J_A^\mu,
\end{equation}
where $J_V^\mu$ is the vector current and $J_A^\mu$ is the
axial-vector current. The matrix element of the vector current in
momentum space of the transition $B \to B' + l^- + \bar{\nu}_l$ is
given by
\begin{eqnarray}
\langle B'(p') \vert &J_V^\mu& \vert B(p) \rangle = \bar{u}'(p') \bigg(
f_1(q^2) \gamma^\mu \nonumber \\
&-& i \frac{f_2(q^2)}{M_B + M_{B'}} \sigma^{\mu \nu}
q_\nu + \frac{f_3(q^2)}{M_B + M_{B'}} q^\mu \bigg) u(p), \nonumber \\
&&
\label{eq:matV}
\end{eqnarray}
where $M_B$ ($M_{B'}$), $p$ ($p'$), $u(p)$ ($u'(p')$), and $\vert B(p)
\rangle$ ($\vert B'(p') \rangle$) are the mass, momentum, Dirac
spinor, and external baryon state of the initial (final) baryon $B$
($B'$), respectively, and $q = p - p'$ is the momentum transfer
\cite{dono92}. The functions $f_i(q^2)$, where $i = 1,2,3$, are the vector
current form factors. These form factors are Lorentz
scalars and contain information about the hadron
dynamics. $f_1$ is the {\it vector} form factor, $f_2$ is the {\it
induced tensor} form factor (or {\it weak magnetism} form factor or
{\it anomalous magnetic moment} form factor), and $f_3$ is the {\it
induced scalar} form factor.

In a previous paper \cite{ohls99}, we have derived the form factors $f_{i}
\equiv f_i(0)$, where $i=1,2,3$,
in the $\chi$QM. Up to terms linear in the parameters
$E \equiv \Delta/\Sigma$ and $\epsilon \equiv \delta/\sigma $,
where $\Sigma = M_B+M_{B'}$, $\Delta = M_B-M_{B'}$, $\sigma =
m_{q}+m_{q'}$, and $\delta = m_{q}-m_{q'}$, these form factors
are
\begin{eqnarray}
f_1 &=& f_1^{\rm QM}, \label{eq:f1app}\\
f_2 &=& \left(\frac{\Sigma}{\sigma} G_A - 1 \right) f_1^{\rm QM},
\label{eq:f2app}\\
f_3 &=& \frac{\Sigma}{\sigma} \left( E G_A - \epsilon
\right) f_1^{\rm QM}.
\label{eq:f3app}
\end{eqnarray}
Here $G_A \equiv g_1^{\rm QM}/f_1^{\rm QM}$, with $f_1^{\rm QM}
\equiv \langle B' \vert \lambda_{qq'} \otimes 1 \vert B \rangle$ and
$g_1^{\rm QM} \equiv \langle B' \vert \lambda_{qq'} \otimes \sigma^z
\vert B \rangle$. The $\lambda_{qq'}$ is the SU(3) matrix that effectuates
the flavor transition and the
$\sigma^z$ operator measures the spin polarizations of the quarks in the
baryons.

\subsection{The weak axial-vector form factors}
\label{sub:GA}

The weak axial-vector form factors $G_A = g_{1}^{\rm QM}/f_{1}^{\rm
QM}$ can be obtained from the SU(6) QM expressed in terms of
the parameters $F$ and $D$
\cite{rent90}. In the $\chi$QM, the
$G_A$'s are expressed in the quark spin polarizations of the proton,
{\it i.e.} $\Delta u$, $\Delta d$, and $\Delta s$. These spin
polarizations differ considerably from the ones in the SU(6) QM due to
the depolarization of the quark spins by the Goldstone bosons (GBs). The
spin polarizations in the $\chi$QM are calculated with one GB emission.
They are \cite{chen95}
\begin{eqnarray}
\Delta u &=& \tfrac{4}{3} - \tfrac{37}{9} a,\\
\Delta d &=& - \tfrac{1}{3} - \tfrac{2}{9} a,\\
\Delta s &=& - a,
\end{eqnarray}
where $a$ is the parameter which measures the probability of emission
of a GB from a quark.
Using the relations $F = \frac{1}{2} (\Delta
u - \Delta s)$ and $D = \frac{1}{2} (\Delta u - 2 \Delta d + \Delta
s)$ \cite{clos93}, we have
\begin{eqnarray}
G_A^{np} &=& \Delta u - \Delta d,\\
G_A^{\Sigma^- \Sigma^0} &=& \tfrac{1}{2} (\Delta u -\Delta s),\\
{g_1^{\rm QM}}^{\Sigma^\pm \Lambda}
&=& \tfrac{1}{\sqrt{6}} (\Delta u - 2 \Delta d + \Delta s),\\
G_A^{\Xi^- \Xi^0} &=& \Delta d - \Delta s
\end{eqnarray}
for the $\Delta S = 0$ decays and
\begin{eqnarray}
G_A^{\Sigma^- n} &=& \Delta d - \Delta s,\\
G_A^{\Xi^- \Sigma^0} &=& \Delta u - \Delta d,\\
G_A^{\Xi^- \Lambda} &=& \tfrac{1}{3} (\Delta u + \Delta d - 2 \Delta
s),\\
G_A^{\Lambda p} &=& \tfrac{1}{3} (2 \Delta u - \Delta d - \Delta s),\\
G_A^{\Xi^0 \Sigma^+} &=& \Delta u - \Delta d
\end{eqnarray}
for the $\Delta S = 1$ decays.

The magnetic moments of the octet baryons and the weak axial-vector
form factor $G_A^{np}$ can be used to fit the parameter $a$ and the
quark magnetic moment $\mu_d$. Using $\mu_u = - 2 \mu_d$ and
$\mu_s = 2 \mu_d/3$ \cite{lind98}, we then obtain $a \simeq 0.104$ and
$\mu_d \simeq -1.196 \, \mu_N$. This gives $\Delta u \simeq 
0.90$, $\Delta d \simeq -0.36$, and $\Delta s \simeq -0.10$.

In the $\chi$QM, the effective quark masses can be determined from the
fitted value of $\mu_d$. The quark masses are then
$m_u^{\rm eff} = m_d^{\rm eff} = m^{\rm eff} \approx 260 \, {\rm MeV}$
and $m_s^{\rm eff} = 3 m^{\rm eff}/ 2\approx 390 \, {\rm MeV}$. In the
following, we will use $m_q \equiv m_q^{\rm eff}$, where $q = u,d,s$.

The values of the $G_A^{BB'}$'s for the $\chi$QM are listed in
Table~\ref{tab:GA}, where for reference also the axial-vector form
factors of the naive QM (NQM) are displayed.

\section{The ratio $\mbox{\boldmath$\rho_f$}$ and the ``weak magnetism''}
\label{sec:ratio}

We will now concentrate on the ``weak magnetism'' form factor
$\rho_f$, which is defined as
\begin{equation}
\rho_f \equiv \frac{f_2}{f_1}.
\label{eq:rhof}
\end{equation}
Inserting Eqs.~(\ref{eq:f1app}) and (\ref{eq:f2app}) in
Eq.~(\ref{eq:rhof}), we obtain
\begin{equation}
\rho_f = \frac{\Sigma}{\sigma} G_A - 1.
\label{eq:rhof1}
\end{equation}
Since there are no linear terms in $E$ and $\epsilon$ in either $f_1$
or $f_2$, the formula for $\rho_f$ is valid up to terms of second
order in $E$ and $\epsilon$.
The quark masses in this and related formulas appear as effective masses,
and the parametric dependence of the quark spin polarization $\Delta
q$, where $q = u,d,s$, on
the emission probability $a$ of GBs incorporates effects of relativistic
corrections and other possible dynamical effects on both the magnetic
moments \cite{dann97} and the $\rho_{f}$'s.  When these effects are taken
into account directly, in terms of a changed structure of the currents, the
fits become worse  \cite{webe97}.

The expression~(\ref{eq:rhof1}) for $\rho_{f}$ above is closely related to the
corresponding formula for the magnetic moments $\mu_B$ of the octet
baryons used in earlier studies. In the same approximation as here, we
have
\begin{eqnarray}
f_{1} &=& Q_B,\\
f_{2} &=& \Sigma \mu_{B} - Q_B =\Sigma \sum_{q=u,d,s}
\frac{e_q}{2m_{q}} \Delta q - Q_B \nonumber \\ &\equiv&
\frac{M_{B}}{M_{N}}\kappa_{B}, 
\label{mag}
\end{eqnarray}
where $e_{q}$ is the quark charge, $Q_B=0, \pm 1$ is the charge of
the baryon, and $\kappa_{B}$ is the anomalous magnetic moment of the
baryon in nuclear magnetons. It is therefore in principle possible to
convert expression~(\ref{eq:rhof1}) above to an expression in terms
of the magnetic moments. This will eliminate the parametric model
dependence. In the following, we will discuss how this can be done in
a way that preserves the absence of terms linear in $E$ and
$\epsilon$ in Eq.~(\ref{eq:rhof1}).

\subsection{The weak magnetism and CVC}

From SU(3) flavor symmetry the weak magnetism form factors can be
related to the magnetic moments of the nucleons. The result is \cite{rent90}
\begin{eqnarray}
\rho^{np}_f & = & (\mu(p)-\mu(n))/\mu_{N}-1,\\
\rho^{\Sigma^{-}\Sigma^{0}}_f & = &
\left(\mu(p)+\tfrac{1}{2}\mu(n)\right)/\mu_{N}-1,\\
f_{2}^{\Sigma^{\pm}\Lambda} & = & -\sqrt{\tfrac{3}{2}}\mu(n)/\mu_{N},\\
\rho^{\Xi^{-}\Xi^{0}}_f & = & (\mu(p)+2\mu(n))/\mu_{N}-1,\\
\rho^{\Sigma^{-}n}_f & = &  (\mu(p)+2\mu(n))/\mu_{N}-1,\\
\rho^{\Xi^{-}\Sigma^{0}}_f & = & (\mu(p)-\mu(n))/\mu_{N}-1, \label{eq:rho_xs}\\
\rho^{\Xi^{-}\Lambda}_f & = & (\mu(p)+\mu(n))/\mu_{N}-1,\\
\rho^{\Lambda p}_f & = & \mu(p)/\mu_{N}-1,\\
\rho^{\Xi^{0}\Sigma^{+}}_f & = & (\mu(p)-\mu(n))/\mu_{N}-1.
\end{eqnarray}
This is called the (extended) CVC hypothesis.
In the NQM, using the formula~(\ref{eq:rhof1}) above, all these relations
emerge by putting $\Sigma/\sigma =M_{N}/m$ and using $\Delta u = 4/3$,
$\Delta d = -1/3$, and $\Delta s = 0.$ Then all $\rho_{f}$'s can be expressed
in terms of the quark magnetic moment $\mu_{d}=-\mu_{u}/2 =-1/(6m)$, which can
be related to the proton and neutron magnetic moments.

It is, however, obvious that the symmetry breaking in the masses, neither
of the quarks nor of the baryons are then accounted for.  In particular,
$\mu_{s}=\mu_{d}$ in this approximation.

On the next level of refinement, one could therefore try to use in
Eq.~(\ref{eq:rhof1}) instead the real baryon masses together with
$m_{s}/m=3/2$,
along with the SU(6) QM values for the spin polarizations $G_{A}$.  Since the
magnetic moments are fairly well accounted for in the NQM, this is probably
a rather good improvement.  In the $\chi$QM, we also allow the spin
polarizations to deviate from their SU(6) values, increasing the
improvement still somewhat.  The results are given in Table~\ref{tab:rho}.

\subsection{The weak magnetism in the chiral quark model. The
$\mbox{\boldmath$\Delta S = 0$}$ cases}

The formula~(\ref{eq:rhof1}) above is transformed
into an expression in terms of the magnetic moments of the baryons,
when
$G_{A}/\sigma$ is expressed in the magnetic moments through $\Delta
q/(2m_{q})$.

Consider for example the $n \to p +
l^- + \bar{\nu}_l$ decay. We can then show, using $\mu(p) = \Delta u \,
\mu_u + \Delta d \, \mu_d +\Delta s \, \mu_s$ and the corresponding
formula for $\mu(n)$, that
\begin{eqnarray}
\rho_f^{n p} &=& \frac{1}{2} \left( 1 + \frac{M_n}{M_p} \right) \left(
\mu(p) - \mu(n) \right) \frac{1}{\mu_N} - 1 \nonumber \\
&\simeq& \left( \mu(p) - \mu(n) \right) \frac{1}{\mu_N} - 1 = \kappa_p -
\kappa_n.
\label{eq:CVC}
\end{eqnarray}
Here we have used the expression $G_A^{n p} = \Delta u - \Delta d$
from Subsection~\ref{sub:GA} above and $\mu_u = - 2
\mu_d$. Equation~(\ref{eq:CVC}) is exactly the CVC formula for the $n
\to p + l^- + \bar{\nu}_l$ decay.

For the other transitions among the octet baryons, the $\chi$QM predicts
the symmetry breaking in these weak magnetic moments due to the symmetry
breaking in the masses both of quarks and baryons. In the following, we
study the symmetry breaking using isospin symmetry and begin
with the $\Delta S=0$ transitions.

For the $\Sigma^{-} \rightarrow \Sigma^{0}$ transition, we have
\begin{equation}
\rho_{f}^{\Sigma^{-}\Sigma^{0}} = \frac{2M_{\Sigma}}{2m} \frac{1}{2}
\left(\Delta u-\Delta s\right) -1.
\label{eq:rho_ss}
\end{equation}
Using the expressions for $\mu(\Sigma^{-})$ and $\mu(\Sigma^{+})$, we
find in this case
\begin{equation}
\Delta u - \Delta s = \frac{1}{3\mu_{d}} \left(\mu(\Sigma^{-}) -
\mu(\Sigma^{+})\right),
\label{eq:rho_ss2}
\end{equation}
which leads to
\begin{eqnarray}
\rho_{f}^{\Sigma^{-}\Sigma^{0}} &=& M_{\Sigma} \left(\mu(\Sigma^{+}) -
\mu(\Sigma^{-})\right) -1 \nonumber \\
&=& \frac{M_{\Sigma}}{2M_{N}}(\kappa_{\Sigma^{+}}-\kappa_{\Sigma^{-}}).
\label{eq:rho_ss3}
\end{eqnarray}
In a similar way, we obtain for the $\Xi^{-} \to \Xi^{0}$ transition
\begin{eqnarray}
\rho^{\Xi^{-}\Xi^{0}}_{f} &=& 2M_{\Xi}
\left(\mu(\Xi^{0})-\mu(\Xi^{-})\right) -1
\nonumber \\
&=& \frac{M_{\Xi}}{M_{N}}(\kappa_{\Xi^{0}}-\kappa_{\Xi^{-}}).
\label{eq:xi1}
\end{eqnarray}

Direct computation in the $\chi$QM using the formula~(\ref{eq:rhof1})
with the parameters $\Delta q$, where $q = u,d,s$, gives
$\rho^{\Xi^{-}\Xi^{0}}_{f} \approx -2.27$, whereas the
formula~(\ref{eq:xi1}) above gives
$\rho^{\Xi^{-}\Xi^{0}}_{f} \approx -1.84$ when the experimental magnetic
moments are inserted. The discrepancy is due to the relatively poor
agreement between the $\chi$QM prediction of the magnetic moments of the
$\Xi^0$ and $\Xi^-$ and their experimental values.

For $f_{1} = 0$, we calculate instead
$f_{2} = \Sigma g_{1}^{QM} / \sigma$. This replaces $\rho_{f}$ for
$\Sigma^{\pm} \to \Lambda$. The result can be expressed in terms of the
$\Sigma\Lambda$
magnetic moment transition matrix element:
\begin{equation}
\mu(\Sigma\Lambda) = - \frac{1}{2\sqrt{3}} \left(\Delta u-2\Delta d + \Delta
s\right)(\mu_{u}-\mu_{d}).
\label{eq:transition_sl}
\end{equation}
We then obtain
\begin{equation}
f_{2}^{\Sigma^{+}\Lambda} = - \sqrt{2} (M_{\Lambda}+M_{\Sigma})
\mu(\Sigma\Lambda).
\label{eq:f2_sl}
\end{equation}

In all the cases above, there is an inherent ambiguity in the choice
of magnetic moments, since in the SU(6) QM the form factors $G_{A}$
can be expressed in only two polarization differences, say $\Delta u-
\Delta d$ and $\Delta u - \Delta s$. In our approximation, the
different choices are related by the sum-rule
\begin{equation}
\mu(p)- \mu(n) +\mu(\Sigma^{-}) - \mu(\Sigma^{+}) +
\mu(\Xi^{0})-\mu(\Xi^{-}) = 0,
\label{eq:sum}
\end{equation}
which follows under quite general assumptions on the spin
polarizations and the magnetic moments of the quarks, and in particular from
the SU(6) QM. This sum-rule is valid to within about $0.5$ $\mu_{N}$ on
the left hand side.

In these cases, $\sigma=2m$ and the quark magnetic moments
$\mu_{u}= -2\mu_{d}=1/(3m)$ are related to $1/\sigma$ without any symmetry
breaking. When we pass to anomalous magnetic moments
in Eqs.~(\ref{eq:CVC}), (\ref{eq:rho_ss3}), and (\ref{eq:xi1}) it is no longer
possible to use Eq.~(\ref{eq:sum}), since extra linear terms in $E$ will
then appear.

\subsection{The $\mbox{\boldmath$\Delta S = 1$}$ cases}

The cases with $\Delta S = 1$ are less straightforward, and there is no
``natural'' way to express the spin polarizations in terms of the
magnetic moments, since many different possibilities give the same
formal result. To begin with, there is a complication that
$\sigma=m+m_{s}$ in these cases. The crucial factor in the
transformation of Eq.~(\ref{eq:rhof1}) into an expression in terms of
magnetic moments is, up to normalization, given by an expression of the form
\begin{equation}
A \simeq \frac{1}{\sigma (\mu_{d}+x\mu_{s})},
\label{A}
\end{equation}
where $x$ is a real parameter. This expression must not contain terms linear in
the small quantity $\epsilon=\delta/\sigma$. It is easy to see that
the condition for this is given by $x=1$.

Let us study this for the case of the $\Sigma^- \to n$ transition. We have
\begin{equation}
\rho_{f}^{\Sigma^- n}=\frac{M_{\Sigma}+
M_{N}}{m+m_{s}}\frac{\mu(\Sigma^{-})-\mu(n)}{\mu_{s}+2\mu_{d}} -1.
\label{sn}
\end{equation}
This can be rewritten as
\begin{equation}
\rho_{f}^{\Sigma^- n}=(M_{\Sigma}+M_{N}) A_{1}(\mu(n)-\mu(\Sigma^{-})) -1,
\label{snb}
\end{equation}
where $A_1 = - 1/((m+m_{s})(\mu_{s}+2\mu_{d}))$. It is
easy to check that this expression is linear in $\epsilon$, since $x =
1/2$. In fact, using $m_{s}=3m/2$, we get $A_1 = 9/10$, so
the deviation from $1$ is $10\%$.

An alternative way of obtaining the spin polarization for the $\Sigma^-
\to n$ transition is to use
\begin{eqnarray}
\mu_{\Sigma n} &\equiv& \mu(n)+\tfrac{1}{2}\mu(p)-\mu(\Sigma^{-})-
\tfrac{1}{2}\mu(\Sigma^{+}) \nonumber \\
&=&-\frac{3}{2} (\Delta d -\Delta s) (\mu_{d}+\mu_{s}).
\label{sn1}
\end{eqnarray}
This gives
\begin{equation}
\rho_{f}^{\Sigma^- n}=(M_{\Sigma}+M_{N}) A_{2} \mu_{\Sigma n} -1,
\label{sn3}
\end{equation}
where
$$
A_{2}= -\frac{2}{3(m+m_{s})(\mu_{s}+\mu_{d})} \simeq 1+ {\cal
O}(\epsilon^{2}).
$$
In fact, for $m_{s}=3m/2$, we obtain $A_{2}=24/25$, which is
only $4\%$ from 1. In the following, this term will therefore be put
equal to $1$. We can then write
\begin{equation}
\rho_{f}^{\Sigma^- n} = \frac{M_{\Sigma}+M_{N}}{2 M_N} \kappa_{\Sigma n},
\label{sn4}
\end{equation}
where $\kappa_{\Sigma n} \equiv
\kappa_{n}+\tfrac{1}{2}\kappa_{p}-\kappa_{\Sigma^{-}} -
\tfrac{1}{2}\kappa_{\Sigma^{+}}$.
Next consider
\begin{eqnarray}
\rho_{f}^{\Xi^{-}\Sigma^{0}} &=& \frac{M_{\Xi}+M_{\Sigma}}{m+m_{s}}(\Delta
u-\Delta d) -1 \nonumber \\
&=& (\mu(p)-\mu(n)) \frac{1}{\mu_{N}} -1 = \kappa_{p}-\kappa_{n},
\label{xs}
\end{eqnarray}
where we have used $M_{\Xi}+M_{\Sigma} \simeq 3(m+m_{s})$. This
expression happens to coincide exactly with CVC (see Eq.~(\ref{eq:rho_xs})).

However, to neglect the hyperfine interaction in the mass formulas
for the baryons means to discard terms linear in the hyperfine
interaction constant, which is generally of the order $50$ MeV. This is
almost of the same order as the symmetry breaking mass difference $\delta$
between the quark masses. We should therefore not be satisfied with this
approximation.

Again, it is possible to use another combination to express the axial-vector
coupling constant. This is given by
\begin{eqnarray}
\mu_{\Xi \Sigma} &\equiv&
\mu(\Sigma^{+})+\tfrac{1}{2}\mu(\Sigma^{-})-\mu(\Xi^{0})-\tfrac{1}{2}
\mu(\Xi^{-}) \nonumber \\
&=& - \frac{3}{2} (\Delta u-\Delta d) (\mu_{d}+\mu_{s}).
\label{eq:mu_xs}
\end{eqnarray}
Using this gives
\begin{equation}
\rho_{f}^{\Xi^{-}\Sigma^{0}} = (M_{\Xi}+M_{\Sigma}) \mu_{\Xi \Sigma}  -1.
\label{xs1}
\end{equation}
For later use this can be rewritten as
\begin{equation}
\rho_{f}^{\Xi^{-}\Sigma^{0}} = \frac{M_{\Xi}+M_{\Sigma}}{2M_{N}}
\left( \kappa_{\Sigma^{+}} + \tfrac{1}{2}\kappa_{\Sigma^{-}} -
\kappa_{\Xi^{0}} - \tfrac{1}{2} \kappa_{\Xi^{-}}\right).
\label{eq:xs2}
\end{equation}

Since all $G_{A}$'s can be expressed in terms of the spin polarization
differences $\Delta d -\Delta s$ and $\Delta u -\Delta d$ it is
possible to express all other $\Delta S=1$ transitions in terms of
$\mu_{\Sigma n}$ and $\mu_{\Xi \Sigma}$. However, if we want to
convert the result from magnetic moments to anomalous magnetic
moments, we must also avoid terms that are linear in the mass ratios
$E=\Delta/\Sigma$.

Consider therefore next
\begin{equation}
\rho_{f}^{\Xi^{-}\Lambda} = \frac{1}{3} \frac{M_{\Xi}+M_{\Lambda}}{m+m_{s}}
\left(\Delta u+\Delta d-2\Delta s \right)-1.
\label{xl1}
\end{equation}
To avoid terms that are linear in $E$, we must in this case use
\begin{eqnarray}
\mu_{\Xi \Lambda} &\equiv& \mu(p)+3\mu(\Lambda) -\mu(\Xi^{0})
-\tfrac{1}{2}\mu(\Sigma^{+})-\tfrac{5}{2}\mu(\Xi^{-}) \nonumber
\\
&=& - \frac{3}{2} \left(\Delta u+\Delta d-2\Delta s
\right)(\mu_{d}+\mu_{s}).
\label{xl3}
\end{eqnarray}
We then obtain, in the same approximation,
\begin{equation}
\rho_{f}^{\Xi^{-}\Lambda} = (M_{\Xi} +M_{\Sigma}) \frac{1}{3} \mu_{\Xi
\Lambda} -1 = \frac{M_{\Xi} +M_{\Sigma}}{2M_{N}} \frac{1}{3} \kappa_{\Xi
\Lambda} + A_3,
\label{eq:xl4}
\end{equation}
where $\kappa_{\Xi \Lambda} \equiv \kappa_{p}+3\kappa_{\Lambda}
-\kappa_{\Xi^{0}}
-\frac{1}{2}\kappa_{\Sigma^{+}}-\frac{5}{2}\kappa_{\Xi^{-} }$ and
$A_{3} =
\frac{1}{3}(M_{\Sigma}+M_{\Xi})(1/(2M_{N})-1/(4M_{\Sigma})+5/(4M_{\Xi}))-1$. 
It is easy to verify that $A_{3}={\cal O}(E^{2})$ and
may therefore be neglected.

Finally, we can express $\rho_{f}^{\Lambda p}$ in terms of magnetic
moments in the same way starting from
\begin{equation}
\rho_{f}^{\Lambda p}= \frac{1}{3} \frac{M_{\Lambda}+M_{N}}{m+m_{s}}
(2\Delta u -\Delta d -\Delta s) -1.
\label{lp}
\end{equation}
The best way of expressing the form factor uses
\begin{eqnarray}
\mu_{\Lambda p} &\equiv& \mu(n)+\tfrac{5}{2}\mu(p)+\tfrac{1}{2}\mu(\Sigma^{-})
-3\mu(\Lambda)- \mu(\Xi^{-}) \nonumber \\
&=& - \frac{3}{2} (2\Delta u-\Delta d-\Delta s)(\mu_{d}+\mu_{s}).
\label{l1}
\end{eqnarray}
Omitting second order mass differences, this gives
\begin{equation}
\rho_{f}^{\Lambda p} = (M_{\Lambda}+M_{N}) \frac{1}{3}\mu_{\Lambda
p} -1 = \frac{M_{\Lambda}+M_{N}}{2M_{N}} \frac{1}{3}\kappa_{\Lambda p},
\label{l2}
\end{equation}
where $\kappa_{\Lambda p} \equiv \kappa_{n}+\frac{5}{2}\kappa_{p}+
\frac{1}{2}\kappa_{\Sigma^{-}}
-3\kappa_{\Lambda}- \kappa_{\Xi^{-}}$.

\subsection{The weak magnetism in the chiral quark soliton model}

The weak magnetic form factors have also  been calculated in the
$\chi$QSM. The result, after normalization in our
convention, is given in \cite{kimp98} as
\begin{eqnarray}
\rho^{np}_f & = & \kappa_{p}-\kappa_{n},\\
\rho^{\Sigma^{-}\Sigma^{0}}_f & = &
\frac{M_{\Sigma}}{2M_{N}}\left(\kappa_{\Sigma^{+}}-
\kappa_{\Sigma^{-}}\right),\\
f_{2}^{\Sigma^{\pm}\Lambda} & = & - \sqrt{2}
\frac{M_{\Sigma}+M_{\Lambda}}{2M_{N}}\frac{\mu_{\Sigma
\Lambda}}{\mu_{N}},\\
\rho^{\Sigma^{-}n}_f & = & \frac{M_{\Sigma}+M_{N}}{2M_{N}} \nonumber \\
&\times&
\left( \kappa_{n} + \tfrac{1}{2}\kappa_{p} - \kappa_{\Sigma^{-}} -
\tfrac{1}{2}\kappa_{\Sigma^{+}} \right),\\
\rho^{\Xi^{-}\Sigma^{0}}_f & = &
\frac{M_{\Xi}+M_{\Sigma}}{2M_{N}} \nonumber \\
&\times& \left( \kappa_{\Sigma^{+}} + \tfrac{1}{2}\kappa_{\Sigma^{-}} -
\kappa_{\Xi^{0}} - \tfrac{1}{2}\kappa_{\Xi^{-}}\right),\\
\rho^{\Xi^{-}\Lambda}_f & = &
\frac{M_{\Xi}+M_{\Lambda}}{2M_{N}}\frac{1}{3} \nonumber \\
&\times& \left(\kappa_{p}+3\kappa_{\Lambda}
-\kappa_{\Xi^{0}}
-\tfrac{1}{2}\kappa_{\Sigma^{+}}-\tfrac{5}{2}\kappa_{\Xi^{-} }\right),\\
\rho^{\Lambda p}_f & = & \frac{M_{\Lambda}+M_{N}}{2M_{N}}\frac{1}{3}
\nonumber \\
&\times& \left( \kappa_{n} + \tfrac{5}{2}\kappa_{p} +
\tfrac{1}{2}\kappa_{\Sigma^{-}} - 3\kappa_{\Lambda}-\kappa_{\Xi^{-}} \right).
\end{eqnarray}
We have here neglected the possible change in the transition from the
anomalous magnetic moments to the full magnetic moments that might be
related to the change in normalization.  By this, we mean that in the above
expressions, $\kappa_{B}=\mu_{B}/\mu_N-Q_{B}M_N/M_{B}$, and the
normalization, of course, is of relevance when the symmetry is broken.

However, our understanding is that, apart from a factor of
$M_B/(M_B + M_{B'})$, these
differences in our normalization are of the order ${\cal O}(m_{s}^{2})$ or
${\cal O}(m_{s}/N_{c}$), {\it i.e.} in terms that are anyhow neglected in
the above formulas \cite{kimp98}, and the corresponding terms of second
order or higher in the mass ratios that are neglected in our calculations.

From Section~\ref{sec:ratio} it should be clear then, that the method of
expressing $\rho_{f}$, that avoids introducing linear terms in the symmetry
breaking masses, in general produces expressions that coincide with those
above.  This shows that the $\chi$QM and the $\chi$QSM give the same
results when linear terms in the symmetry breaking are eliminated.

\section{Discussion}

The particular choice of combinations of magnetic
moments, that enables one to express the $\rho_{f}$'s in term of
anomalous magnetic moments, are enforced from the cancellation of linear
terms in both the quark and baryonic mass differences.
The analysis presented here shows that when this is done 
the $\chi$QM and the $\chi$QSM give the same results.
The earlier noticed numerical differences are related to the
difficulty to reproduce the octet baryon magnetic moments in the
$\chi$QM without symmetry breaking in the spin polarizations. This can
be seen {\it e.g.} in the case of $\Xi^0$ and $\Xi^-$.

It is of course possible to stop and be satisfied at the level where
the $\rho_{f}$'s
are expressed in terms of magnetic moments. Then, since all $G_{A}$'s
can be expressed in terms of only two spin polarization differences,
there are several equivalent relations for the $\rho_{f}$'s related to sum
rules for the
$G_{A}$'s. On top of that, there is in this case also the possibility to
use the sum-rules for the magnetic moments to find alternative ways to express
the $\rho_{f}$'s.

For the $\rho_f$'s the existing experimental data is given in
Table~\ref{tab:rho}.

Let us consider this table.

The CVC values listed are in a
way half experimental results, since they use the measured values of
the anomalous magnetic moments for the nucleons as input data to
calculate these values.

Since the magnetic moments of the quarks are fitted to the
magnetic moments of the proton and neutron, the SU(6) QM results
should coincide with CVC in the absence of symmetry breaking and are
not listed.

All values obtained for the $\rho_f$'s in the $\chi$QM lie within the
experimental errors, where experimental data exist. (The experimental
results have large errors, though.)

In one case, that of neutron decay, we can see that $\rho_{f}(\chi{\rm
QM}) \approx \rho_{f}({\rm CVC})$. For the other decays, the
$\rho_{f}$'s of the $\chi$QM incorporate effects of vector current
non-conservation due to the mass differences between the
isomultiplets as well as depolarization of the spin due to GB emission.

All calculated values for the $\chi$QM have
the same sign as the CVC values and they are also close in
magnitude. 
The numerical results cannot be expected to be much better than within
$10\%$. Already isospin is violated to a few percent.

For comparison, we have in two cases calculated the $\rho_{f}$'s obtained by
neglecting the hyperfine interaction, since it is rather small. The
results are 
$$
\rho_{f}^{\Xi^{-}\Sigma^0} = (\mu(p) - \mu(n))/\mu_N - 1 \simeq 3.71
$$
and
\begin{eqnarray}
\rho_{f}^{\Xi^{-}\Lambda} &=& (\mu(\Sigma^{+})
+\mu(\Xi^{0}) -\mu(\Xi^{-}) - \mu(\Sigma^{-}))/(3\mu_{N}) -1 \nonumber \\
&=& 0.01 \pm 0.02, \nonumber
\end{eqnarray}
which both are very close to the values using
Eqs.~(\ref{eq:xs2}) and (\ref{eq:xl4}), respectively.

\section{Summary and conclusions}

We have studied the baryonic weak magnetism form factors in
detail in the spirit of the $\chi$QM and compared the results with
the $\chi$QSM. The comparison shows that the results are in good
agreement, and that the differences are of the order of reliability
of the results in all cases. This might indicate that the main part
of the symmetry breaking is accounted for in these formulas.
The numerical results are presented in Tables~\ref{tab:GA} and \ref{tab:rho}.

The present investigation has used the SU(3) symmetric coupling in the
$\chi$QM and the static approximation for the quarks. A natural
improvement would be to incorporate
lowest order non-static effects and further SU(3) symmetry breaking
effects \cite{chen98,song97}, to obtain better agreement with
experimental data. In particular, we expect that this would lead to a
closer agreement with the $\rho_f$ ratios obtained from direct
application of Eq.~(\ref{eq:rhof1}), since symmetry breaking can better account
for the
octet baryon magnetic moments \cite{lind98}. SU(3) symmetry breaking
also leads to better agreement for $g_A^{n p}$
\cite{lind98,chen98,song97}.

\begin{acknowledgement}
{\it Acknowledgments.} This work was supported by the Swedish Natural Science
Research Council (NFR), Contract No. F-AA/FU03281-312. Support for
this work was also provided by the Engineer Ernst Johnson Foundation (T.O.).
\end{acknowledgement}

\begin{threeparttable}
\caption{Weak axial-vector form factors, $G_A^{BB'}$. The values in
the NQM column are the SU(6) values for the weak axial-vector form
factors and the values in the $\chi$QM column are obtained from the
quark spin polarizations.
${g_1^{\rm QM}}^{\Sigma^\pm \Lambda}$ are given instead of $G_A^{\Sigma^\pm
\Lambda}$, since ${f_1^{\rm QM}}^{\Sigma^\pm \Lambda} = 0$. The
experimental values for $G_{A}^{BB'}$ have been 
obtained from Ref.~\protect\cite{caso98}, except for the ${{g_1^{\rm
QM}}^{\Sigma^- \Lambda}}$ and ${G_A^{\Xi^- \Sigma^0}}$ values, which
are CERN WA2 \protect\cite{bour82,gail84} results from branching
ratio measurements}
\begin{tabular}{lcrr}
\hline
\\
Quantity & Experimental & NQM & $\chi$QM\\
& value\\
\hline
$G_A^{np}$ & $1.2670 \pm 0.0035$ & $\frac{5}{3}$ & $1.26$\\
$G_A^{\Sigma^- \Sigma^0}$ & - & $\frac{2}{3}$ & $0.50$\\
${g_1^{\rm QM}}^{\Sigma^- \Lambda}$ & $0.589 \pm 0.016$ &
$\sqrt{\frac{2}{3}}$ & $0.62$\\
${g_1^{\rm QM}}^{\Sigma^+ \Lambda}$ & - & $\sqrt{\frac{2}{3}}$ &
$0.62$\\
$G_A^{\Xi^- \Xi^0}$ & - & $-\frac{1}{3}$ & $-0.25$\\
\hline
$G_A^{\Sigma^- n}$ & $-0.340 \pm 0.017$ & $-\frac{1}{3}$ & $-0.25$\\
$G_A^{\Xi^- \Sigma^0}$ & $1.25 \pm 0.15$ & $\frac{5}{3}$ & $1.26$\\
$G_A^{\Xi^- \Lambda}$ & $0.25 \pm 0.05$ & $\frac{1}{3}$ & $0.25$\\
$G_A^{\Lambda p}$ & $0.718 \pm 0.015$ & $1$ & $0.76$\\
$G_A^{\Xi^0 \Sigma^+}$ & - & $\frac{5}{3}$ & $1.26$\\
\hline
\end{tabular}
\label{tab:GA}
\end{threeparttable}

\begin{threeparttable}
\caption{The ratios $\rho_f^{BB'} \equiv
\frac{f_2^{BB'}}{f_1^{BB'}}$. The experimental values have been
obtained from
Ref.~\protect\cite{bour82} (see also Ref.~\protect\cite{gail84}).
$f_2^{\Sigma^\pm \Lambda}$ are given instead of $\rho_f^{\Sigma^\pm
\Lambda}$, since $f_1^{\Sigma^\pm \Lambda} = 0$}
\begin{tabular}{lcccc}
\hline
\\
Quantity & Experimental & CVC & $\chi$QM & $\chi$QSM \cite{kimp98} \\
 & value\\
\hline
$\rho_f^{np}$ & $3.71 \pm 0.00$ (input) & 3.71 & 3.53 & 3.71\\
$\rho_f^{\Sigma^- \Sigma^0}$ & - & 0.84 & 1.31 & 1.30\\
$f_2^{\Sigma^- \Lambda}$ & $3.52 \pm 3.52$ & 2.34 & 2.73 & 2.80\\
$f_2^{\Sigma^+ \Lambda}$ & - & 2.34 & 2.72 & 2.80\\
$\rho_f^{\Xi^- \Xi^0}$ & - & $-2.03$ & $-2.27$ & -\\
\hline
$\rho_f^{\Sigma^- n}$ & $-1.78 \pm 0.61$ & $-2.03$ & $-1.82$ & $-1.67$\\
$\rho_f^{\Xi^- \Sigma^0}$ & - & 3.71 & $3.85$ & $3.61$\\
$\rho_f^{\Xi^- \Lambda}$ & $-0.44 \pm 0.46$ & $-0.12$ & $-0.06$ & $0.10$\\
$\rho_f^{\Lambda p}$ & $2.43 \pm 1.49$ & 1.79 & $1.38$ & $1.52$\\
$\rho_f^{\Xi^0 \Sigma^+}$ & - & 3.71 & $3.83$ & -\\
\hline
\end{tabular}
\label{tab:rho}
\end{threeparttable}

\end{document}